\documentclass[11pt,a4paper]{article}
\usepackage{amsmath}
\usepackage{latexsym}
\usepackage[all]{xy}
\usepackage{theorem}
\newtheorem{teorema}{Theorem}[section]

\include{amslatex}
\newtheorem{proposicion}[teorema]{Proposition}

{\theorembodyfont{\rmfamily}
}
{\theorembodyfont{\rmfamily} }

\numberwithin{equation}{section}

\include{amsmath}

\begin{document}
\begin{title}
{\LARGE {\bf Averaged Lorentz Dynamics and an application in Plasma Dynamics}}
\end{title}
\maketitle
\author{
\begin{center}
Ricardo Gallego
Torrome\footnote{r.gallegotorrome@lancaster.ac.uk$\,\,\,$
Supported by EPSRC and Cockcroft Institute}

Department of Physics, Lancaster University,
\\
Lancaster, LA1 4YB \& The Cockcroft Institute WA4 4AD, UK\\[3pt]

\end{center}}
\bigskip
\

\begin{abstract}
Using a geometric averaging procedure applied to a non-affine linear connection, we prove that for a narrow one particle distribution function and in the ultra-relativistic limit, a bunch of charged point particles can be described by a Charged Cold Fluid Model, without additional hypothesis on the moments.
\end{abstract}
\bigskip
\bigskip

\section{Introduction}
Modeling the dynamics of a non-neutral plasma is an important problem in Plasma Physics and its applications in Beam Dynamics. Each bunch of a beam in an accelerator contains about $10^{10}$ particles in a small region. Therefore, one looks for models such that: 1. Are simple enough to be useful in numerical simulations of the dynamics and 2. Contains the main features of the particle nature of the constitutive elements of the bunches.

The traditional approach has been to use fluid models. Since the bunches are formed by charged particles, a natural view is to interpret the fluid model as a macroscopic approximation of a kinetic model [1]. The resulting model is simple and still capable to retain some particle features coming from the underlying Kinetic Theory.
However, these derivation is based on some assumptions that maybe are not accomplished in modern accelerators. These assumptions are usually on the high order moments of the distribution function. This seems to be a general feature of all the derivations of fluid models from Kinetic Theory [1].

We present in this note a new derivation of the cold fluid model from Kinetic Theory which uses only natural hypothesis, happening in current particle accelerator machines. The way to do this is the following. Firstly, we re-write the Lorentz force equation as an auto-parallel condition of a non-affine linear connection. Then, we use an averaging procedure described in [2] to average this connection. The resulting averaged connection is an affine connection on the manifold {\bf M}. After this, one compares the dynamics of both connections [3]. For narrow distributions which follow the Vlasov equation and in the ultra-relativistic dynamics, both dynamics are similar [3].
 Therefore, we have two Kinetic Models (one based on the Vlasov equation and the other on the Liouville equation associated with the averaged connection). It happens that under the same assumptions than for the particle dynamics, the corresponding solutions of the Vlasov equation $f$ and the averaged Vlasov equation $\tilde{f}$ are similar [4]. One can also prove that the corresponding velocity vector fields are similar. Finally, the velocity field of the Liouville equation associated with the averaged dynamics is controlled by the diameter of the distribution and by the energy of the bunch [3]. Considering all together, one has control of how good is the cold fluid model as an approximation of the Vlasov's model.

{\bf Notation}. The space-time manifold {\bf M} is $n$-dimensional. The metric $\eta$ is the Minkowski metric, the potential $A$ is a smooth $1$-form.
\section{The Lorentz Connection}

The Lorentz force can be written in a covariant form in a general coordinate system in the following way:
\begin{equation}
\frac{d^2 \sigma^i}{dt^2} +\, ^{\eta}\Gamma^i\,_{jk} \frac{d
\sigma^j}{dt}\frac{d \sigma^k}{dt} +\eta^{ij}(dA)_{jk} \frac{d
\sigma^k}{dt}\sqrt{\eta(\frac{d \sigma}{dt},\frac{d
\sigma}{dt})}=0,\quad i,j,k=0,1,...,n,
\end{equation}
where $\sigma: {\bf I}\longrightarrow {\bf M}$ is a solution curve for $t\in {\bf I}$, $^{\eta}\Gamma^i\,_{jk}$ are the coefficients of the Levi-Civita connection $^{\eta}\nabla$ of $\eta$ and $dA$ is the exterior derivative of the $1$-form $A$ (electromagnetic potential); here the parameter $t$ is the proper-time of $\eta$ of the curve $\sigma$.

This system of second order differential equations are an auto-parallel condition of a non-affine linear Koszul connection [5]:
\begin{equation}
^L\nabla_{\dot{\sigma}} \dot{\sigma}=0.
\end{equation}
The connection coefficients are
\begin{displaymath}
^L\Gamma^i\,_{jk}(x,y)=\, ^{\eta}\Gamma^i\,_{jk} +
\frac{1}{2{\sqrt{\eta(y,y)}}}({\bf F}^i\,_{j}(x)y^m\eta_{mk}+ {\bf
F}^i\,_{k}(x)y^m\eta_{mj})+
\end{displaymath}
\begin{equation} +{\bf F}^i\,_m (x)
\frac{y^m}{2\sqrt{\eta(y,y)}}(\eta_{jk}-\frac{1}{\eta(y,y)}\eta_{js}
\eta_{kl}y^s y^l),
\end{equation}
with ${\bf F}_{ij}:=\partial_i A_j -\partial_j A_i$ and ${\bf F}^i\,_j=\eta^{ik}{\bf F}_{kj}$.
This defines a non-affine connection on the pull-back bundle defined by the following diagram:
\begin{equation}
\xymatrix{\pi^* {\bf TM} \ar[d]_{\pi_1} \ar[r]^{\pi_2} &
{\bf TM} \ar[d]^{\tilde{\pi}}\\
{\bf N}  \ar[r]^{\pi} & {\bf M},}
\end{equation}
where
\begin{displaymath}
{\bf N}:=\bigsqcup_{{\bf x\in {\bf M}}} \,\{y\in \,{\bf T}_x{\bf M},\,|\,\eta_{ij}(x)y^i\,y^j>0\}.
\end{displaymath}
\section{The averaged Lorentz dynamics}

The Lorentz connection $^L\nabla$ is not an affine connection on {\bf M}. One way to obtain an affine connection is to integrate over the support of the distribution $f$ the connection coefficients [3]. This construction has a geometric interpretation [2], but is not relevant here. Applying this procedure to eq. (1.2), one obtains that the averaged connection $<\,^L\nabla>$ has the following coefficients:
\begin{displaymath}
<\,^L\Gamma^i\,_{jk} >=\, ^{\eta}\Gamma^i\,_{jk}+ ({\bf
F}^i\,_{j}<\frac{1}{2\sqrt{\eta(y,y)}}y^m>\eta_{mk}+ {\bf
F}^i\,_{k}<\frac{1}{2\sqrt{\eta(y,y)}}y^m>\eta_{mj})+
\end{displaymath}
\begin{equation}
+{\bf F}^i\, _m\,\frac{1}{2}\big(
<\frac{y^m}{({\eta(y,y)})^{3/2}}>\,\eta_{jk}-\eta_{js}
\eta_{kl}<\frac{1}{(\eta(y,y))^{3/2}}\,y^m y^s y^l>\,\big).
\end{equation}
Each of the integrations is equal to the $y$-integration along the
fiber:
\begin{displaymath}
<y^i> :=\frac{1}{vol({\bf \Sigma}_x)}\int _{{\bf \Sigma}_x} y^i
f(x,y)\,d\mu,\quad {vol({\bf \Sigma}_x)}=\int _{{\bf \Sigma}_x}
f(x,y)\,d\mu
\end{displaymath}
and similarly for higher moments. Here ${\bf \Sigma}$ is the unit hyperboloid, defined by
\begin{displaymath}
{\bf \Sigma}:=\bigsqcup_{x\in {\bf M}}\,\{y\in\,{\bf T}_x{\bf M}\,|\,\eta_{ij}(x)y^i y^j=1\,\}.
\end{displaymath}
${\bf \Sigma}_x$ is the restriction of ${\bf \Sigma}$ to ${\bf T}_x{\bf M}$.
\begin{proposicion} Let ${\bf M}$ and $^L\nabla$ be as before. Assume that the support of the distribution $f:{\bf \Sigma}\longrightarrow {\bf R}$ is compact and denote by $<\,^L\nabla>$ the averaged Lorentz connection. Then:
\begin{enumerate}

\item The connection $<\,^L\nabla>$ is an affine, symmetric connection on ${\bf
M}$. Therefore, for any point $x\in {\bf M}$, there is a {\it normal coordinate system} such that the averaged coefficients are
zero.

\item To write down the form of $<\,^L\nabla>$ we only need the first, second and third moments of the distribution function $f(x,y)$.
\end{enumerate}
\end{proposicion}
\section{Comparison between $^L\nabla$ and $<\,^L\nabla>$}

If on ${\bf  M  }$ there is a time-like vector field $U$ normalized such that
$\eta(U,U)=1$, one can define the Riemannian metric $\bar{\eta}$:
\begin{equation}
\bar{\eta}(X,Y):=-\eta(X,Y)+2\eta(X,U)\eta(Y,U).
\end{equation}
Using $\bar{\eta}$ there is a scalar product on the
vector space ${\bf T}_x{\bf M}$ defined by
$\bar{\eta}(x)_{ij} dy^i\otimes dy^j$. The diameter of the distribution $f_x$ is
\begin{displaymath}
\alpha_x:=sup\{d_{\bar{\eta}}(y_1,{y}_2)\, |\, y_1,{y}_2\in
sup(f_x )\},\,\,\,{\alpha}:=sup\{{\alpha}_x,\, x\in {\bf M}\}.
\end{displaymath}
The Riemannian metric $\bar{\eta}$ induces a distance
function $d_{\bar{\eta}}$ on the manifold ${\bf T}_x{\bf M}$.
Then, we define
${\alpha}:=sup\{{\alpha}_x,\, x\in {\bf M}\}$.
We choose as vector field $U$ in the definition of the Riemannian metric $(6.1)$ the following:
\begin{equation}
U(x)=\frac{<\hat{y}>(x)}{\eta_{ij}(x)\,<\hat{y}^i>(x)<\hat{y}^j>(x)>},\, \,\textrm{if}\,\,\eta_{ij}(x)<\hat{y}^i>(x)<\hat{y}^j>(x)\,> 0
\end{equation}
and $0$ if $\eta_{ij}(x)<\hat{y}^i>(x)<\hat{y}^j>(x)\,\leq 0$. Given a continuous operator $A_x:{\bf T}_x{\bf
M}\longrightarrow {\bf T}_x{\bf M}$, its operator norm is
\begin{displaymath}
\|A\|_{\bar{\eta}}(x):=sup\big\{\,\frac{\|A(y)\|_{\bar{\eta}}}{\|y\|_{\bar{\eta}}} (x),\,y\in{\bf T}_x{\bf M}\setminus \{0\}\,\big\}.
\end{displaymath}
Let us denote by $\bar{\gamma}(t)$ the gamma factor of the Lorentz transformation from the local frame defined by the vector field $U$ to the laboratory frame, at some instance defined by the laboratory local time coordinate $t$. Denote by $\theta ^2(x)=\vec{y}^2(x)-\,<\vec{\hat{y}}>^2(x)$ and $\bar{\theta}^2(x)=<\vec{\hat{y}}>^2(x)-\vec{\tilde{y}}^2(x)$. Here $\vec{y}(t)$ is the velocity tangent vector field along a solution of the Lorentz force equation and $\tilde{y}(t)$ is spatial component of the tangent vector field along a solution of the averaged equation, with both solutions having the same initial conditions. The maximal values of this quantities on the compact space-time manifold are denoted by $\theta^2$ and $\bar{\theta}^2$.
Then we can stay the following theorems [3]:
\begin{teorema}
Let ${\bf M}$ be a semi-Randers space such that $\eta$ is the Minkowski metric. Let us assume that:
\begin{enumerate}

\item The auto-parallel curves of unit velocity of the connections $^L\nabla$ and $<\,^L\nabla>$ are defined for values of laboratory frame coordinate time at least $t$.

\item The dynamics occurs in the ultra-relativistic limit, $E(x)>>1$ for all $x\in {\bf M}$.

\item The distribution function is narrow, $\infty >\,E(x)>>{\alpha}$ for all $x\in {\bf M}$.

\item It holds the following inequality:
\begin{displaymath}
|\theta^2\,-\bar{\theta}^2|\ll 1
\end{displaymath}
\item The support of the distribution function $f$ is invariant under the flow of the Lorentz force equation
\end{enumerate}
Then, for the same arbitrary initial condition, the solutions of the equations
\begin{displaymath}
^L\nabla_{\dot{x}} \dot{x}=0,\, \quad <\,^L\nabla>_{\dot{\tilde{x}}} \dot{\tilde{x}}=0
\end{displaymath}
 are such that:
\begin{equation}
\|\tilde{x}(t)-\, x(t)\|\leq\, \big(C(x)\|{\bf F}\|(x)\,+C^2_2(x)(1+B_2(x){\alpha})\big){\alpha}^2\,E^{-2}(x)\,t^2,
\end{equation}
where $C(x)$, $C_2(x)$ and $B_2(x)$ are functions on {\bf M} and bounded by constants of order $1$.
\end{teorema}
\begin{teorema}
Under the same hypothesis as in {\it theorem
4.1}, the difference between the tangent vectors is given by
\begin{equation}
\|\dot{\tilde{x}}^i(t)-\dot{x}^i(t)\|\leq \big(K(x)\|{\bf F}\|(x)\,+K^2_2(1+D_2(x){\alpha})\big){\alpha}^2\,E^{-2}\,t.
\end{equation}
with ${K}_i$ and $D_2(x)$ functions bounded by constants of order 1.
\end{teorema}

\section{Results on Kinetic Theory and Fluid Models}
In this section we overview our results from [4] on the approximation of the Vlasov Kinetic Model by Cold Fluid Model.

In Kinetic Theory, the $1$-particle distribution function follows
the Liouville equation:
\begin{equation}
\chi f(x,y)=0,
\end{equation}
where $\chi$ is the {\it Liouville vector field}. In the case that the non-linear connection is
the Lorentz connection, we consider the Vlasov equation:
\begin{equation}
^L\chi f(x,y) =0,
\end{equation}
with the vector field
\begin{displaymath}
^L\chi=y^k \partial_k
-\,^L\Gamma^k\,_{ij}y^iy^j\frac{\partial}{\partial y^k},\quad i,j,k=0,1,...,n.
\end{displaymath}
Similarly, for the averaged connection, the corresponding
Liouville's equation is given by
\begin{equation}
<\,^L\chi> f(x,y) =0,
\end{equation}
with the vector field
\begin{displaymath}
\quad <\,^L\chi>:=y^k \partial_k
-<\,^L\Gamma^k\,_{ij}>y^iy^j\frac{\partial}{\partial y^k},\quad i,j,k=0,1,...,n.
\end{displaymath}

This is a partial integro-differential equation where the averaged vector field $<\,^L\chi> $ is obtained using
the distribution function associated with the Liouville
vector field of the averaged connection $<\,^L\nabla>$. However, it is easier
(because the existence of normal coordinates [5]) to extract some consequence from this equation. In particular,
\begin{proposicion}
Let $f$ and $\tilde{f}$ be solutions of the Liouville equations $^L\chi (f) =0$ and $<\,^L\chi
> (\tilde{f}) =0$ such that
 $\tilde{\alpha}$, the diameter of $\tilde{f}$ is small; $^L\chi$ and $<\,^L\chi
>$ are the spray vector fields obtained from the connections $^L\nabla$ and $<\, ^L\nabla>$.
Assume that
$\tilde{\alpha}<<E=\,<y^0>_{\tilde{f}}$ in the
laboratory frame and the averaged is performed using $\tilde{f}$.
Then for the solutions of the Vlasov and the {\it averaged Vlasov equation},
\begin{displaymath}
|f(x(t),y(t))-\tilde{f}(x(t),y(t))|<C_{\bf M}\cdot |x(t)-\tilde{x}(t)|
\end{displaymath}
for some constant $C_{\bf M}$ depending on the manifold {\bf M}.
\end{proposicion}
One also proves the following [4]:
\begin{proposicion}
With the above notation, the following relation holds:
\begin{equation}
<\, ^L \chi> \tilde{f}=0 \,\,\Rightarrow \,\,<\,
^L\nabla>_{\tilde{V}} \tilde{V}=\,O(\tilde{\alpha}^2),
\end{equation}
with $\tilde{\alpha}$ the diameter of the distribution $\tilde{f}$.
\end{proposicion}

The natural implication of the above results is the following [4]:

\begin{teorema}
Let {\bf F} be an external electromagnetic field and $^L\nabla$
the associated Lorentz connection. Then the solutions of
 the Lorentz force equation
 \begin{equation}
^L\nabla_{\dot{x}} {\dot{x}}= 0
 \end{equation}
can be approximated by the integral curves of the normalized mean velocity field $u$ of the
distribution function $f(x,y)$, solution of the associated Vlasov equation
$^L\chi f=0$ and the difference is controlled by polynomials on $\tilde{\alpha}$ at least of order $2$.
\end{teorema}

This theorem is the main result presented in this paper. It affirms the possibility of
 using the cold fluid model for narrow distributions and in the limit ultra-relativistic limit.

\end{document}